# Energy-efficient population coding constrains network size of a neuronal array system

Lianchun Yu

Institute of Theoretical Physics, Lanzhou University, Lanzhou, 730000, China

Key Laboratory for Magnetism and Magnetic Materials of the Ministry of Education,

Lanzhou University, Lanzhou, 730000, China

State Key Laboratory of Theoretical Physics, Institute of Theoretical Physics, Chinese

Academy of Sciences, Beijing, 100109, China

Chi Zhang

Cuiying Honors College, lanzhou University, Lanzhou, 730000, China

Liwei Liu

College of Electrical Engineering, Northwest University for Nationalities, Lanzhou, 730070,

China

Yuguo Yu*

State Key Laboratory of Medical Neurobiology, Center for Computational Systems Biology,

School of Life Sciences, Fudan University, Shanghai, 200433, China

Department of Neurobiology, Yale University School of Medicine, New Haven, CT 06520,

USA

**Correspondence:** yuguo.yu@yale.edu

**Abstract**

Here, we consider the open issue of how the energy efficiency of neural information transmission process in a general neuronal array constrains the network size, and how well this network size ensures the neural information being transmitted reliably in a noisy environment. By direct mathematical analysis, we have obtained general solutions proving that there exists an optimal neuronal number in the network with which the average coding energy cost (defined as energy consumption divided by mutual information) per neuron passes through a global minimum for both subthreshold and superthreshold signals. Varying with increases in background noise intensity, the optimal neuronal number decreases for subthreshold and increases for suprathreshold signals. The existence of an optimal neuronal number in an array network reveals a general rule for population coding stating that the neuronal number should be large enough to ensure reliable information transmission robust to the noisy environment but small enough to minimize energy cost.

**INTRODUCTION**

Neuronal activity related to information processing in brain circuits is metabolically expensive[1]. For example, the metabolic cost of the human brain may rise from 20% to 40% of whole-body energy production when human beings switch from a resting state to a working state. Action potentials, which are electrical signals and rely on the potential energy stored in transmembrane ion gradients, cost a large fraction of this energy[2]. These metabolic demands could be large enough to influence the design, function and evolution of brains[3-10].

Population coding[11], i.e., the cooperative coding of information of input signals by a group of neurons, is a basic neural code strategy used in many nervous systems[12]. Studies have shown that neuronal activities could be synchronized to remain robust against noise perturbation and promote reliable information transmission[13-15]. It was suggested[16,17] that a certain number of neurons involved in the synchronous neuronal activities is critical to deliver information reliably within a feed-forward multi-layer cortical network. However, for population coding in such an array network, how energy is efficiently consumed related to information transmission and neuronal number has not been carefully considered, especially in the case of different background noise levels.

Considering that the metabolic consumption of an individual neuron is almost ten times the consumption of a body cell, the number of neurons involved in a neuronal circuit during an information-processing task may dominate the energy cost of the system. Moreover, in reality, background noise is present at all levels of the nervous system, from the microscopic level, such as channel noise in membranes and biochemical noise at synapses, to macroscopic levels[18-20]. The existence of noise may degrade the reliability of effective information transmission and requires the involvement of more neurons to perform an information-processing task[15]. Therefore, it will be critical to address the issue of what is the proper size of a neuronal array network for reliable information transmission with minimal energy cost in a noisy environment.

In an earlier time, Barlow (1961)[21] suggested sparseness as one of the principles important to sensory representation. Because sparse codes are defined as representations with low activity ratios—i.e., at any given time a small proportion of neurons are active—they are sometimes proposed as a means to help conserve metabolic costs. Levy and Baxter (1996)[22] demonstrated that there should exist an optimal firing probability for any given mean firing rate in a neural network so that sensory information representation capacity to energy expenditure is maximized. Later a number of studies[5,7-10,23-25] among the pioneers have linked the energy efficiency of spiking generation process with the number of the ion channels[19], as well as the ratio of sodium to potassium channel density within a single neuron[5,7,9,23-25]. However, these pioneering works did not consider well the situation of existence of different noise level in a neuronal network on the neural code capacity and energy expenditure. Especially, in the past several decades, a bunch of studies on stochastic resonance had shown that noise play important roles in neural information processing for sub- or suprathreshold signals[14,26,27]. Therefore, it is still an open issue to study the energy efficiency of a group of neurons with dynamical model to understand its dependence on the population size, signal and noise intensity.

Here, we first solved a stochastic one-dimensional bistable Langevin equation, which mimics the action potential generations with a particle crossing the barrier of a double well, to obtain an analytical solution for the pulse signal detection rate and spontaneous firing rate[28]. Coincidence detector (CD)[29-32] in the context of neurobiology is a process by which a neuron or a neural circuit can encode information by detecting the occurrence of temporally close but spatially distributed input signals from presynaptic neurons of a network. A bunch of reports[31-33] suggested that neuronal network with postsynaptic CD might be popular in different cortical areas to read synchronous activities in the noisy background. Hence, here we constructed an array network model with $N$ bistable neurons and a CD neuron to pool up the network information, and calculated the mutual information and energy cost of the network.

**Results**

The bistable neuron model used here can be described with the following equation:

$$\dot{v} = -U'(v) + \Gamma(t) \tag{1}$$

where $v$ is the membrane potential, and $U$ is a double well potential, defined as:

$$U = -\frac{a}{2}v^2 + \frac{v^4}{4} \tag{2}$$

Note that $U$ has two minima at $v_{s1} = -\sqrt{a}$, $v_{s2} = \sqrt{a}$ and a saddle point at $v_\mu = 0$. In the following calculation, $a = 1$ by default. $\Gamma(t)$ is the Gaussian white noise, with

$$<\Gamma(t)>=0; <\Gamma(t)\Gamma(t')>=2D\delta(t-t') \tag{3}$$

where D is the noise intensity. We assume the neuron to be at its resting state when the particle is in the left well and excited when the particle crosses the barrier to the right well due to noise perturbation or signal stimulation.

It is assumed that a force with a short time duration moves the particle horizontally from the resting state to $v'$ in the region of the saddle point. When the force disappears, the particle drifts up to the region of the saddle point. Near the saddle point, the phase trajectories are repelled, causing the particle to accelerate away from the saddle-point region towards one of two minima. According to Lecar and Nossal's approach of linearizing around the saddle point, we can obtain the probability of finding the particle in the right well after a long enough time, i.e., the probability that a pulse input signal is detected by the neuron[20,34],

$$P_c(\Delta v) = \frac{1}{2}[1 + erf(\frac{\Delta v}{\sqrt{2D/a}})], \quad (4)$$

where $\Delta v = v' - v_u$ is the strength of the pulse inputs, and $erf(x) = \frac{2}{\sqrt{\pi}} \int_0^x \exp(-t^2)dt$ is the Gaussian error function. With no input signal, neurons fire spikes spontaneously under noise perturbation. The spontaneous firing rate of the bistable neuron is derived by Kramers' formula[35]:

$$P_s = \frac{\sqrt{2}a}{2\pi} \exp(-\frac{a^2}{4D}). \quad (5)$$

Figure 1(a) shows the firing probability of the neuron as a function of input pulse strength $\Delta v$ with different noise intensities D, described by Eq.4. It is clear that the firing threshold fluctuates depending on the strength of noise perturbation. The neuron fires in response to subthreshold inputs ($\Delta v < 0$) with the assistance of noise, which is well known as stochastic resonance[28]. The detection rate increases as the noise intensity increases. However, the noise sabotages the neuron's reliability for suprathreshold inputs ($\Delta v > 0$), and the detection rate decreases as the noise intensity increases. For threshold inputs ($\Delta v = 0$), the detection rate is 50%, independent of noise intensity. Our results are consistent with previous simulation results for a Hodgkin-Huxley (HH) system (for example, see Fig. 4 in reference [36] in which the inputs are in the form of voltage pulses). Fig.1(b) shows that the spontaneous firing rate increases as a function of noise intensity. The same result was obtained by Zeng et al. with the stochastic simulation of the HH neuron[37].

Next we considered the information capacity and energy cost of an array of $N$ bistable neurons with a pulse signal as input whose intensity is distributed uniformly over the interval $[\Delta v_{min}, \Delta v_{max}]$, with the probability density function $p(\Delta v) = \begin{cases} \frac{1}{\Delta v_{max} - \Delta v_{min}} & \Delta v_{min} \leq \Delta v \leq \Delta v_{max} \\ 0 & \text{otherwise} \end{cases}$, and the mean input strength $\bar{v} = \frac{\Delta v_{max} + \Delta v_{min}}{2}$. The information in the synchronous firings of this N neuron array is pulled together by a CD neuron, see Fig. 1(c). The CD neuron receives the outputs of the $N$ bistable neuron array and is excited if n ( $\theta \leq n \leq N$ ) inputs arrive simultaneously, where $\theta$ is the threshold of the CD neuron. In response to pulse inputs, this network has two output values, i.e., *R= {r| r=1 if CD neuron fires, or =0 if CD neuron fails to fire}*. The conditional probability $q(r=1|\Delta v)$ that the CD neuron fires when the input is $\Delta v$ is given by a cumulative binomial distribution

$$q(r=1|\Delta v) = \sum_{k=\theta}^{N} \binom{N}{k} (P_c(\Delta v))^k \cdot (1-P_c(\Delta v))^{N-k}, \quad (6)$$

where $\binom{N}{K}$ is the binomial coefficient, and $P_c(\Delta v)$ is the detection rate of the bistable neuron for pulse input with strength $\Delta v$, determined by Eq. 4. Then, the conditional probability that the CD neuron does not fire when the input is $\Delta v$ is given by

$$q(r=0|\Delta v) = 1 - q(r=1|\Delta v). \quad (7)$$

According to the Bayes formula, the probability that the output is $r$ can be obtained by

$$q(r \in R) = \int_S q(\Delta v) q(r|\Delta v) d(\Delta v). \quad (8)$$

According to Shannon's information theory[38], the information between input $S$ and output $R$ is defined as

$$I_M(S;R) = \sum_{\Delta v \in S} \sum_{r \in R} p(\Delta v) q(r | \Delta v) \log_2 \frac{q(r | \Delta v)}{q(r)}. \quad (9)$$

In our case, the input is continuous and the output is discrete, and thus, the summation must be rewritten as follows:

$$I_M(S;R) = \sum_{r \in R} \int_S p(\Delta v) q(r | \Delta v) \log_2 \frac{q(\mathrm{r} | \Delta v)}{q(\mathrm{r})} d(\Delta v). \quad (10)$$

Finally, we obtain the following description of the mutual information for the CD neuron:

$$I_M(S;R) = \frac{1}{\Delta v_{max} - \Delta v_{min}} \sum_{i=0,1} \int_{\Delta v_{min}}^{\Delta v_{max}} q(r = i | \Delta v) \log_2 \frac{q(\mathrm{r} = i | \Delta v)}{q(\mathrm{r} = i)} d(\Delta v). \quad (11)$$

Fig.2(b) shows that $I_{single} = I_M / N$, the average mutual information per single neuron can reach a global maximum when the network contains an optimal number of neurons for subthreshold signals (e.g., $\Delta v_{sub} = -0.1$). The optimal neuronal number becomes smaller when the noise intensity increases. For superthreshold signals (e.g., $\Delta v_{sub} = 0.1$), the average mutual information per single neuron can also be maximized by an optimal neuronal number. However, when the noise intensity decreases, the optimal neuronal number also decreases (see Fig.2(b)). From Fig.1(a), we see that for both subthreshold and superthreshold signals, $P_c$ converges to 0.5 as the noise intensity increases. As a result, in Fig.2(a) and (b), the optimal neuronal number moves to 20 with the CD threshold set to 10(results not shown). In contrast, with decreasing noise, to re-establish a maximum, the subthreshold signals require an increased number of neurons to compensate for the loss due to decreased $P_c$, whereas the suprathreshold signals must decrease the number of neurons to limit the excess due to decreased $P_c$. For both the subthreshold and superthreshold signals, the average mutual information per neuron can be maximized by an optimal noise intensity, displaying a classic subthreshold stochastic resonance phenomenon (see Fig.2(c)) and suprathreshold stochastic resonance phenomenon (see

Fig.2(d))[39].

For an *N* neuron array system, the total network energy expenditure in an action potential onset time interval $\Delta t$ can be written as

$$E_{\Delta t} = E_s(N)\Delta t + \int_S d(\Delta v) p(\Delta v) E_{\Delta v}(N,t), \tag{12}$$

where $E_s(N)$ and $E_{\Delta v}(N,t)$ are the energy costs related to action potential generation. For simplicity, we assume the energy cost of one action potential to be 1. $E_s(N)$ is the energy cost of the spontaneous firings in unit time, and $E_s(N) = NP_s$. $E_{\Delta v}(N,t)$ is the energy cost of the action potentials in response to input pulses with strength $\Delta v$, and $E_{\Delta v}(N,t) = NP_c(\Delta v)$ if the inputs are applied at the beginning of this time interval and zero otherwise. Therefore, $\int_S d(\Delta v) p(\Delta v) E_{\Delta v}(N,t)$ is the average energy cost of action potentials in response to an input pulse with distribution $p(\Delta v)$. Fig. 3 (a) shows the dependence of $E_{single} = E_{\Delta t}/\Delta t N$, the average energy cost of each neuron in unit time, as a function of input pulse strength $\Delta v$ for different noise intensities. Note that when the noise is weak, as the spikes are mostly induced by the signals, the $E_{single} - \bar{v}$ curves have similar behavior to the $P_c - \bar{v}$ curves shown in Fig. 1(a). Interestingly, for subthreshold signals, e.g., $\Delta v_{sub}$=-0.1, $E_{single}$ increases as noise intensity D increases, while for superthreshold signals, e.g., $\Delta v_{supra}$=0.1, $E_{single}$ first decreases and then increases slightly as noise intensity D increases (see Fig. 3(b)). For subthreshold signals, e.g., $\Delta v_{sub}$=-0.1, $E_{system} = E_{\Delta t}/\Delta t$, the energy consumption of the whole system in unit time, increases monotonously with both noise intensity and neuronal number; see Fig.3(c). For superthreshold signals, e.g., $\Delta v_{supra}$=0.1, $E_{system}$ also increases with neuronal number. However, $E_{system}$ is large for weak noise intensity and becomes small for high noise intensity, see Fig.3(d).

We now define a new measurement to quantify how efficiently the system utilizes a certain amount of energy in a certain amount of information transmission, i.e., energy cost per unit information transmission or coding energy cost: $\eta = \frac{E}{I}$, where E is the average energy consumption in unit time per neuron, and I is the average mutual information between the inputs and outputs of the neuron array in unit time per neuron. Now, we have

$$\eta = \frac{E_s(N)\Delta t + \int_S d(\Delta v) p(\Delta v) N P_C(\Delta v)}{I_M}. \qquad (13)$$

Laughlin et al. found that in noise-limited signaling systems, a low capacity weak pathway transmits information more economically, which promotes the idea of distributed information coding among multiple pathways [3]. The analysis result for the array network supports this idea. Fig.4(a) shows that for subthreshold signals, though the network detection rate for a single neuron is low, it yields a low coding energy cost in the information coding process for weak noise intensity. Moreover, our results show that the coding energy cost passes through a global minimum as a function of neuronal number within the network for different noise intensities. The optimal neuronal number $N_{opt}$, corresponding to the minimum coding energy cost $\eta_{opt}$, shifts to the smaller number as noise intensity increases. For a suprathreshold stimulus, the coding energy cost also passes through a global minimum as a function of neuronal number. However, the optimal neuronal number corresponding to the minimum coding energy cost shifts to the larger number side as noise intensity increases (see Fig.4(b)). Interestingly, as the noise intensity increases, the optimal neuronal number $N_{opt}$ for both sub- and suprathreshold signals converges to a small range between N=15 and 25 (see Fig.4(c)). This convergence occurs because the optimal neuronal number for maximal mutual information, as we discussed above (Fig.2(a) and (b)), will converge from the opposite direction to 20 in the case of the large noise limit, recalling that the energy cost does not change greatly for different noise intensities.

Moreover, we found that for a given noise intensity (e.g., D=0.5), the maximal input pulse frequency that the bistable neurons can receive, which is the inverse of the action potential onset time $\Delta t$, can significantly modulate the values of either $N_{opt}$ or $\eta_{opt}$ for different input pulse intensities, suggesting an energy saving mechanism for information coding in higher frequency bands, as observed in recent experimental findings[25].

**Discussion**

Consuming only several watts of energy, mammalian brains are able to carry out 1000 trillion operations per second. The biophysical mechanism of this extremely efficient energy expenditure is still not fully known. In a real living brain circuit, background noise is present at all levels of the nervous system, from the microscopic level, such as channel noise in membranes and biochemical noise at synapses, to macroscopic levels, such as a small neuronal circuit composed of several to tens of neurons. The existence of noise may degrade the reliability of effective information transmission and requires the involvement of more neurons to perform an information-processing task[15]. For example, small neurons will cost less energy because fewer ion channels are involved, thus requiring less ion exchange through the ion pumps that drive ATPase Na+/K+ exchanges after action potentials[9]. However, the stochastic nature of ion channel gating will not only produce variability in the response of neuron to external stimuli but also cause spontaneous action potentials, damaging the reliability of signal processing[18]. In this case, trade-offs between information transfer and energy cost may constrain the proper number of ionic channels in individual neurons[19,20] as well as the proper size of neuronal number in a neuronal network. Considering that the metabolic consumption of an individual neuron is almost ten times the consumption of a body cell, the number of neurons involved in a neuronal circuit during an information-processing task may dominate the energy cost of the system. Therefore, there may exist a general rule for energy consumption, neural information transmission and network size.

In this paper, we have examined the energy efficiency for an information coding process based on a neuronal array network composed of $N$ simple bistable neurons and a CD detector neuron. We have provided an analytical solution that quantifies the relationships among the energy cost per neuron, mutual information, noise intensity, signal intensity and frequency, and neuronal number required in the circuit for effective information transmission. The novel result obtained here is to reveal a general rule for energetics related to population coding that there exists an optimal number of neurons in the network necessary for maximal information coding with minimal energy cost. The optimum depends on the noise intensity, input pulse strength and frequency. The results reflect general mechanisms for sensory coding processes, which may give insight into energy efficient brain communication and neural coding.

Historically, Barlow introduced the concept of the economy of impulses to argue through evolution, the neuronal codes should minimize its representational entropy and use lower and lower levels of cell firings to produce equivalent encoding [29]. Levy and Baxter factored energy expenditures into the economy of impulses to demonstrate that for both binary and analog neurons, there exists an optimal firing rate probability for maximal information transmission with minimal energy cost [30]. Our work may among the first to reveal a general principle that there always exists an optimal neuronal number with which the information coding capacity of the network could be maximized while the energy cost is relatively low in the presence of different noise levels. We carried out mathematical analysis on a simple neuronal model here to make the analysis solution general applied for all kinds of excitable neurons. However, this model is not complex enough to include one detail of ionic channel properties contributing to energy efficient spiking process while a lot of recent research focus on this issue and found[5,7-9,23,25] that there may exist an optimal number of ionic channels[19] or appropriate ratio of sodium to potassium channel density to general energy efficient action potentials[5,9,23,25]. Our model rather considered a more general situation of a network case in the presence of noise situation, and proved that the coding energy cost per unit information goes through a global minimum with optimal neuronal number depending on given noise intensity.

In our work, introducing of the bistable model makes it possible to analyze directly the input-output response function to the signals in noise environment, and provide a general solution about energy-constrained efficient information transmission process. Since bistable state describes the action potential initiation process in HH systems, our results presented here can be applied to those HH-type models, and explain the realistic biological issues. In addition, although our work here focused on the effects of system size on energy efficiency, it could be extended to include the effects of spike correlation, stimulus and noise distribution on the energy efficiency, based on the recent progress on suprathreshold stochastic resonance[14,27,40]. It is worth to mention that our analysis considered in detail the contribution of noise to information transmission regarding sub- and super-threshold stochastic resonance, while this could not be derived from pure binary neurons in the early studies[22].

**Acknowledgement**

This work was supported by the National Natural Science Foundation of China (Grant No. 11105062) and the Fundamental Research Funds for the Central Universities (Grant No. lzujbky-2011-57 and No. lzujbky-2015-119). Dr. Yu also offers thanks for the support from the National Natural Science Foundation of China (31271170) and the program for the Professor of Special Appointment (Eastern Scholar SHH1140004) at Shanghai Institutions of Higher Learning.

**Additional Information:** The authors declare no competing financial interests.

**Figure Legends**

Fig. 1: (a) The detection rate of the bistable neuron model as a function of input pulse strength for different noise intensities. (b) The bistable neuron model's spontaneous firing rate as a function of noise intensity (b). (c) The network model with an array of N neurons and a coincidence detector (CD) neuron with a spiking threshold $\theta = 10$.

Fig.2: The average mutual information per neuron $I_{single}$ as a function of array size N for subthreshold signal $\Delta v_{sub}$=-0.1 (a) and for suprathreshold signal $\Delta v_{supra}$=0.1 (b). (c) $I_{single}$ vs. noise intensity D for neuronal numbers N=20, 30, and 40. (d) $I_{single}$ vs. D for N=10, 15, and 20.

Fig.3:(a) The average energy cost per neuron $E_{single}$ as a function of input pulse intensity $\Delta v$ for different noise intensities D. (b) $E_{single}$ as a function of D for $\Delta v_{sub}$=-0.1 and $\Delta v_{supra}$=0.1, respectively. (c) The total network energy consumption $E_{system}$ as a function of neuronal number N for $\Delta v_{sub}$=-0.1. (d) $E_{system}$ as a function of neuronal number N for $\Delta v_{supra}$=0.1.

Fig. 4: (a) Coding energy cost η as a function of N for $\Delta v_{sub}$ =-0.1 in the cases of different noise intensities. (b) η vs. N for $\Delta v_{supra}$ =0.1 in the cases of different noise intensities. (c) The optimal neuronal number $N_{opt}$ vs. noise intensity D for different input signal intensities $\Delta v$. (d) The minimum coding energy cost $\eta_{opt}$ vs. the optimal neuronal number $N_{opt}$ for different signal intensities $\Delta v$ in the case of different input pulse frequencies.

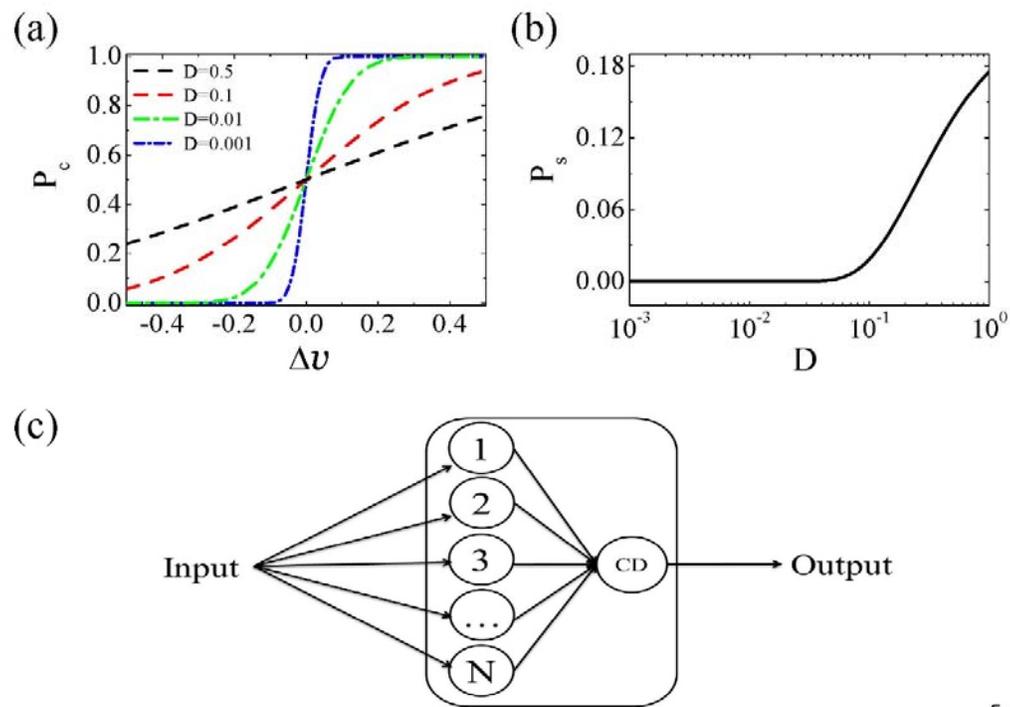

fig.1

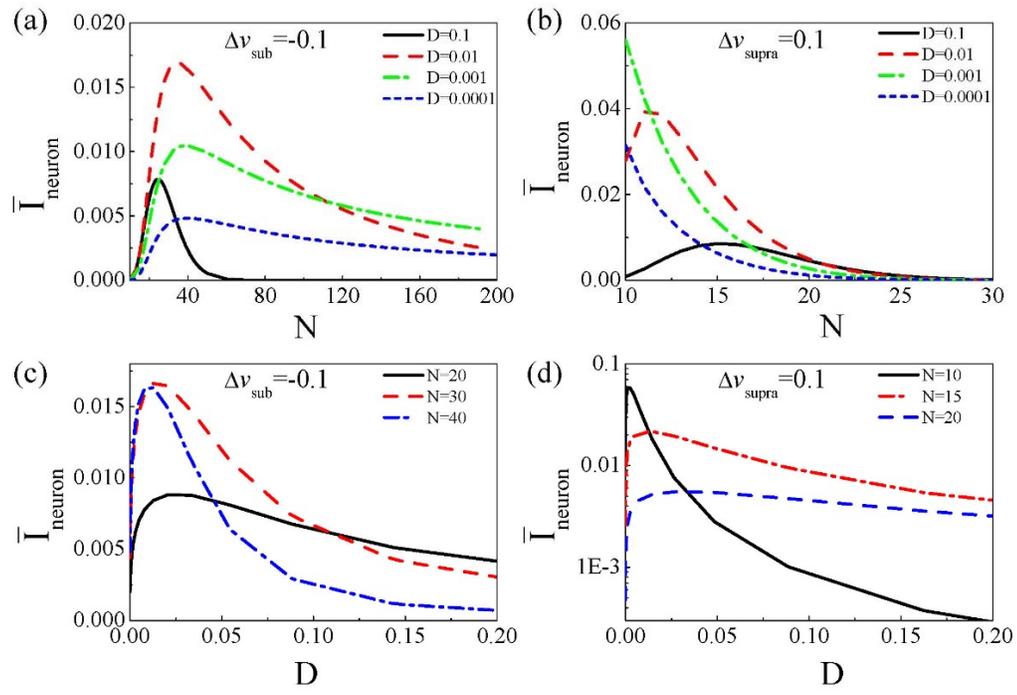

N for $\Delta v_{sub}$=-0.1. (d) $E_{system}$ as a function of neuronal number N for $\Delta v_{supra}$=0.1.

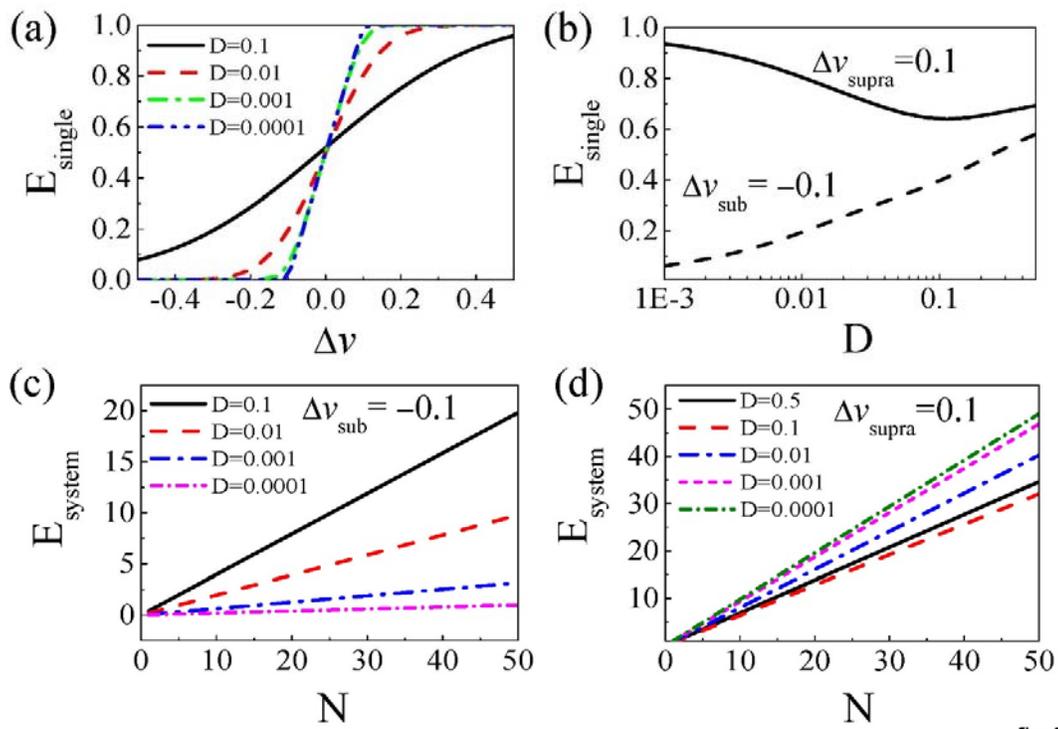

fig.3

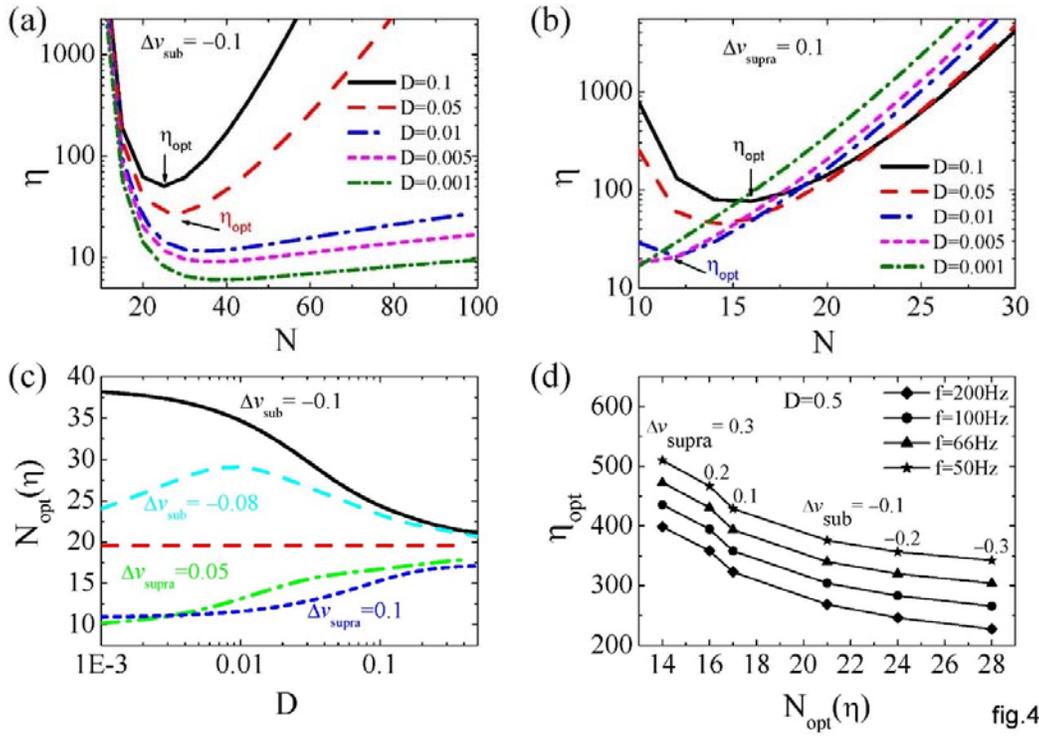